\documentclass[eqsecnum,aps,nofootinbib]{revtex4}

\usepackage{graphicx}
\usepackage{bm}
\usepackage{theorem}
\usepackage{here}

\begin{document}

\title{Cosmological Co-evolution of Yang-Mills Fields and Perfect Fluids}

\date{\today }

\author{John D. Barrow$^{1}$} 
\author{Yoshida Jin$^{2}$}
\author{Kei-ichi Maeda$^{2,3,4}$}
\affiliation{$^{1}$DAMTP, Centre for Mathematical Sciences, 
Cambridge University, Cambridge CB3 0WA, UK}
\affiliation{$^{2}$Department of Physics, Waseda University, 
3-4-1 Okubo, Shinjuku-ku,Tokyo 169-8555, Japan}
\affiliation{$^{3}$ Waseda Institute for Astrophysics, Waseda University,
Shinjuku, Tokyo 169-8555, Japan}
\affiliation{$^{4}$ Advanced Research Institute for Science and Engineering,
Waseda  University, Shinjuku, Tokyo 169-8555, Japan}

\begin{abstract}
We study the co-evolution of Yang-Mills fields and perfect fluids in Bianchi
type I universes. We investigate numerically the evolution of the universe
and the Yang-Mills fields during the radiation and dust eras of a universe
that is almost isotropic. 
The Yang-Mills field undergoes small
amplitude chaotic oscillations, 
which are also displayed by the expansion scale factors of the universe. 
The results of the numerical simulations are 
interpreted analytically and compared with past studies of the cosmological
evolution of magnetic fields in radiation and dust universes. We find that,
whereas magnetic universes are strongly constrained by the microwave
background anisotropy, Yang-Mills universes are principally constrained by
primordial nucleosynthesis and the bound  is comparatively weak, and 
$\Omega _{\mathrm{YM}}<0.105\Omega _{\mathrm{rad}} \,.$
\end{abstract}

\maketitle

\section{Introduction}

There is considerable interest in the generation and evolution of magnetic
fields in cosmological models \cite{sub}. This interest has a double focus.
There is the hope that an explanation might be found for the existence of
significant magnetic fields in galaxies. This may require seed fields to
originate very early in the history of the universe, or even form part of
the initial conditions. We have also discovered that the cosmological
evolution of anisotropic stresses, like electric and magnetic fields, has
interesting mathematical features that create unexpected physical
consequences. These features appear when a homogeneous and anisotropic
stress with a trace-free energy-momentum tensor is present during the
radiation era of the universe. The isotropic black-body radiation and the
anisotropic stresses evolve to first order in the same way, but there is a
greatly slowed (logarithmic) decay of the shear anisotropy caused by the
pressure anisotropy of the anisotropic fluid \cite{zeld},\cite{collins}. If
the evolution were to be linearised around the isotropic model then a zero
eigenvalue would exist. When the equation of state of the accompanying
perfect fluid changes from radiation to dust the evolution of the shear is
still dominated by the pressure anisotropy of the anisotropic stresses but
the zero eigenvalue disappears and the shear falls as a reduced power of the
cosmic time. In the study of these effects made in ref \cite{Barrow} we
considered the effects of anisotropic stresses whose trace-free anisotropic
pressure tensor $\pi _{ab}$ was proportional to a density $\mu _{A},$ so

\begin{equation}
\pi _{ab}=C_{ab}\mu _{A},  \label{an}
\end{equation}%
with $C_{ab}$ constant and $C_{a}^{a}=0.$ Barrow and Maartens considered the
generalisation of these results to general inhomogeneous cosmological models
close to isotropy using the covariant formalism \cite{BM} and also some
applications to Kaluza-Klein cosmologies \cite{BMKK}. The key feature in the
evolution of almost-isotropic cosmological models containing perfect fluids
plus a magnetic field is that during the radiation era the ratio of the
shear to Hubble expansion rate falls only logarithmically in time. The
universe evolves towards an attractor state in which it is proportional to
the ratio of the magnetic and perfect-fluid densities. This means that the
ratio of these densities at the epoch of last-scattering determines the
large-angle temperature anisotropy of the cosmic microwave background (CMB).
This enables us to use observations of the CMB to place strong bounds on any
homogeneous cosmological magnetic field \cite{maglim}, or other anisotropic
stress defined by a constant $C_{ab},$ \cite{Barrow}. An important extension
of these studies is to consider anisotropic stresses with more general
pressure tensors, for example those with time-dependent $C_{ab}.$

An important case with time-dependent $C_{ab}$, which generalises the
electromagnetic field, is that of a Yang-Mills stress. This has been studied
for pure Yang-Mills fields and no accompanying fluid by several authors in
simple Bianchi type I anisotropic universes \cite{YMtypeI, kun, blev}, in
all class A Bianchi type universes \cite{jin}, and in Kantowski-Sachs
universes \cite{YMKS}. These studies reveal that the Yang-Mills field
creates a form of chaotic behaviour during the evolution of the Yang-Mills
stress \cite{blev}. This is not surprising because such behaviour is present
in Minkowski space-time and is nothing to do with the cosmological evolution
or the spacetime curvature. It should not be confused with the chaotic
behaviour of general relativistic origin that is found in the 'Mixmaster'
universes of Bianchi types VIII and IX, even in vacuum\footnote{%
The most general cosmological analysis of the Yang-Mills chaos is that
carried out for all Ellis-MacCallum Class A Bianchi-type vacuum cosmologies by Jin
and Maeda \cite{jin} and they are able to consider the simultaneous presence
of the Yang-Mills chaos and Mixmaster chaos \cite{mix, bkl, jdbchaos1,
jdbchaos2, jdbchaos3}.}. 
Therefore the Yang-Mills chaos occurs in the
simplest axisymmetric universes of Bianchi type I and is present even when
the anisotropy level is arbitrarily small. This raises the interesting question of
whether the chaotic evolution might leave an imprint in the temperature
anisotropy of the CMB when it decouples at a redshift $z_{rec}\approx 1000.$

However, all of these earlier studies consider the evolution of
Einstein-Yang-Mills equations only for the case of a pure Yang-Mills stress
in an anisotropic cosmological model. Our experience \cite{Barrow} with the
behaviour of magnetic fields in anisotropic universes teaches us that it is
important to include the presence of a perfect fluid in order to find the
realistic evolution of the Yang-Mills stress during the radiation and
dust-dominated eras of the cosmological expansion. In this paper we consider
this generalisation by analysing the evolution of spatially homogeneous,
anisotropic universes of Bianchi type I containing both a perfect fluid and
an anisotropic Yang-Mills stress. By combining analytic and numerical
studies we determine the evolution of the shear and the Yang-Mills stress
during the early universe in order to discover if it is possible for it to
leave an observable signature in the CMB.

In section 2 we set out the Einstein-Yang-Mills equations for the Bianchi
class A universes. In section 3 we specialise to consider in detail the
evolution for the Bianchi type I universe containing Yang-Mills fields and
perfect fluids, focussing on the physically significant cases of
pressureless dust and black-body radiation. In section 4 we describe the
numerical and analytical results and in section 5 compare them in detail
with the situation that results when a pure magnetic field replaces the
Yang-Mills field. We consider the observational bounds that can be placed on
the magnetic and Yang-Mills fields in the presence of perfect fluids using
the microwave background temperature anisotropy and the primordial
nucleosynthesis of helium-4 in section 6. We summarise our results in
section 7.

\section{The Einstein-Yang-Mills evolution equations}

The Einstein equations to be solved for universes containing a perfect fluid
and a Yang-Mills (YM) field are 
\begin{eqnarray}
G_{\mu \nu } &=&8\pi GT_{\mu \nu }, \\
T_{\mu \nu } &=&T_{\mathrm{YM}\mu \nu }+T_{\mathrm{m}\mu \nu }, \\
T_{\mathrm{m}\mu \nu } &=&\mbox{diag}
  (-\mu _{\mathrm{m}},p_{\mathrm{m}},p_{\mathrm{m}},p_{\mathrm{m}}), \\
T_{\mathrm{YM}\mu \nu } &=&\frac{1}{g_{\mathrm{YM}}^{2}}\left[ F_{\mu 
\lambda }^{(A)}F_{\ \nu }^{(A)\ \lambda }-\frac{1}{4}g_{\mu \nu }F_{\lambda 
\sigma }^{(A)}F^{(A)\lambda \sigma }\right] ,
\end{eqnarray}%
where $F_{\mu \nu }^{(A)}$ is the field strength of YM field,
\textquotedblleft $A$\textquotedblright\ describes the components of the
internal SU(2) space, and $\mu _{\mathrm{m}}$ and $p_{\mathrm{m}}$ have 
an equation of state

\begin{center}
$p_{\mathrm{m}}=(\gamma -1)\mu _{\mathrm{m}}.$
\end{center}

We shall assume that $T_{\mathrm{YM}\mu \nu }$ and $T_{\mathrm{m}\mu \nu }$
are separately conserved and there is no energy exchange between them.
Hence, from the vanishing covariant divergence of $T_{\mathrm{YM}\mu \nu }$
we find that the YM equations are 
\begin{equation}
\vec{F}_{\ \ \ ;\nu }^{\mu \nu }-\vec{A}_{\nu }\times \vec{F}^{\mu \nu }=0\,,
\end{equation}%
where $g_{\mathrm{YM}}$ is the self-coupling constant of YM field, and $\vec{ 
F}^{\mu \nu }=F^{(A)\mu \nu }\vec{\tau}_{A}$ and $\vec{A}_{\nu }=A_{\nu
}^{(A)}\vec{\tau}_{A}$ with $\vec{\tau}_{A}$ being the $SU(2)$ basis.
Setting our units as $8\pi G/g_{\mathrm{YM}}^{2}=1$ \footnote{%
This choice of units has the following consequences. If $g_{\mathrm{YM}}=1$,
then $8\pi G=1$; if $g_{\mathrm{YM}}=10^{-10}$, then $8\pi G=10^{-20}$ and
in this case the unit of time is $10^{10}t_{\mathrm{pl}}\sim 10^{-34}\mathrm{ 
sec}$ and the present Hubble time is $H_{0}^{-1}=1.2\times 10^{10}\mathrm{yr} 
=4\times 10^{17}\mathrm{sec}=10^{51}$.} 
and $c=\hbar =1$, the basic
equations become free of the value of $g_{\mathrm{YM}}$.

We adopt the orthonormal-frame formalism, which has been developed in ref. 
\cite{wainwright-ellis}. In Bianchi-type spacetimes, there exists a
3-dimensional homogeneous spacelike hypersurface $\Sigma _{t}$ which is
parameterized by a time coordinate $t$. The timelike basis is given by 
$\bm{e}_{0}=\bm{\partial}_{t}$. The triad basis $\{\bm{e}_{a}\}$ on the 
hypersurface $\Sigma _{t}$ is defined by the commutation function $\gamma
_{\ ab}^{c}$ as 
\begin{equation}
\lbrack \bm{e}_{a},\ \bm{e}_{b}]=\gamma _{\ ab}^{c}\bm{e}_{c}\,.
\end{equation}%
It is convenient to decompose $\gamma _{\ ab}^{c}$ into the geometric
variables denoted conventionally by the Hubble expansion $H$, the
acceleration $\dot{u}_{a}$, the shear $\sigma _{ab}$, and the vorticity $%
\omega _{ab}$, and the variables $\Omega _{a}$ (the rotation of $\bm{e}_{a}$%
), and the variables $a_{a}$ and $n_{ab}$, which distinguish the type of
Bianchi model, so 
\begin{eqnarray}
\gamma _{\ 0b}^{a} &=&-\sigma _{b}^{\ a}-H\delta _{b}^{\ a}-\epsilon _{\
bc}^{a}(\omega ^{c}+\Omega ^{c}),  \label{gamma1} \\
\gamma _{\ 0a}^{0} &=&\dot{u}_{a},  \label{gamma2} \\
\gamma _{\ ab}^{0} &=&-2\epsilon _{ab}^{\quad c}\omega _{c},  \label{gamma3}
\\
\gamma _{\ ab}^{c} &=&\epsilon _{abd}n^{dc}+a_{a}\delta _{b}^{\
c}-a_{b}\delta _{a}^{\ c}.  \label{gamma4}
\end{eqnarray}%
\emph{\ }If Bianchi spacetimes are expressed in this way, the vorticity $\omega
_{ab}$ always vanishes because they are hypersurface orthogonal.

In order to analyze the EYM system, we shall study the simplest case. First,
using the gauge freedom of the YM field, we set $A_{0}^{(A)}(t)=0$, which
simplifies the vector potential to $\vec{\bm{A}}=A_{a}^{(A)}(t)\vec{\tau}_{A} 
\bm{\omega}^{a}$, where $\bm{\omega}^{\alpha }$ is the dual basis of $\bm{e} 
_{\alpha }$ \cite{okuyama}. Next we shall restrict attention to cosmologies
of Bianchi type class A, in which the vector $a_{a}$ vanishes. We can also
diagonalize $n_{ab}$, i.e. $n_{ab}=\mbox{diag}\{n_{1},n_{2},n_{3}\}$, using
the remaining freedoms of a time-dependent rotation of the triad basis.
Then, we can show that if $\sigma _{ab},\ A_{a}^{(A)}$ and $\dot{A} 
_{a}^{(A)} $ do not have off-diagonal components initially, the equations of
motion guarantee that those variables will remain diagonal during the
subsequent evolution for the class A Bianchi spacetimes (see the Appendix in 
\cite{jin}). As a result, $\Omega _{a}$ vanishes and the number of basic
variables defining the initial value problem reduces from 21 (12 for
spacetime [$H,N_{a},\sigma _{ab},\Omega _{a}$] and 9 for YM field 
[$A_{a}^{(A)}$]) to 10 (7 for spacetime [$H,N_{a},\sigma _{aa}$] and 3 
for YM field [$Y_{a}^{(a)}$]).

In order to discuss the dynamics most efficiently, it is convenient to
introduce the Hubble-normalized variables, which are defined as follows: 
\begin{equation}
\Sigma _{ab}\equiv \frac{\sigma _{ab}}{H},~~~\ N_{a}\equiv \frac{n_{a}}{H}.
\label{def}
\end{equation}

We also re-express the shear variables of the Bianchi spacetime as 
\[
\Sigma _{+}\equiv \frac{1}{2}\left( \Sigma _{22}+\Sigma _{33}\right)
,~~\Sigma _{-}\equiv \frac{1}{2\sqrt{3}}\left( \Sigma _{22}-\Sigma
_{33}\right) ,~~\Sigma ^{2}\equiv \Sigma _{+}^{2}+\Sigma _{-}^{2}\,. 
\]%
The diagonal components of the YM field potential are described by new
variables $a(t),\ b(t),$ and $c(t)$ as 
\[
a\equiv A_{1}^{(1)},\ b\equiv A_{2}^{(2)},\ c\equiv A_{3}^{(3)}\,. 
\]%
and a new time variable, $\tau $, defined by $d\tau =Hdt$ is introduced so
that $\tau $ denotes the e-folding number of the scale length.

Using these variables we can write down the evolution and constraint
equations explicitly. They consist of the generalized Friedmann equation,
the dynamical Einstein equations and the YM evolution equations. The
generalized Friedmann equation, which is the constraint equation, is 
\begin{equation}
\Sigma ^{2}+\Omega _{\mathrm{YM}}+\Omega _{\mathrm{m}}+K=1\,,
\label{eq:GFRW}
\end{equation}%
where $\Omega _{\mathrm{YM}}$ is the density parameter of the YM field, i.e.
the Hubble-normalized energy density of the YM field, which is defined by 
\begin{eqnarray}
\Omega _{\mathrm{YM}} &\equiv &\frac{\mu _{\mathrm{YM}}}{3H^{2}}  \nonumber
\\
&=&\frac{1}{6}\left[ \left\{ a^{\prime }+(-2\Sigma _{+}+1)a\right\}
^{2}+\left\{ b^{\prime }+(\Sigma _{+}+\sqrt{3}\Sigma _{-}+1)b\right\}
^{2}+\left\{ c^{\prime }+(\Sigma _{+}-\sqrt{3}\Sigma _{-}+1)c\right\}
^{2}\right.  \nonumber \\
&&\left. +\left( N_{1}a+{\frac{bc}{H}}\right) ^{2}+\left( N_{2}b+{\frac{ca}{H%
}}\right) ^{2}+\left( N_{3}c+{\frac{ab}{H}}\right) ^{2}\right] ;
\end{eqnarray}%
$\Omega _{\mathrm{m}}$ is the density parameter of perfect fluid, defined by 
$\Omega _{\mathrm{m}}\equiv \mu _{\mathrm{m}}/(3H^{2})$, and $K$ is the
Hubble normalized curvature, which is defined by 
\[
K\equiv -\frac{{}^{(3)}R}{6H^{2}}=\frac{1}{12}\left\{
N_{1}^{2}+N_{2}^{2}+N_{3}^{2}-2(N_{1}N_{2}+N_{2}N_{3}+N_{3}N_{1})\right\}
\,. 
\]%
Note that the positive 3-curvature corresponds to $K<0$. Hence, except for
the Bianchi type IX models we have $K\geq 0$. The energy density of YM field
and perfect fluid are always positive definite. Thus we find that $\Sigma
^{2},\ \Omega _{\mathrm{YM}},$ and $K$ are restricted to the domains $0\leq
\Sigma ^{2}\leq 1,\ 0\leq \Omega _{\mathrm{YM}}\leq 1,$ and $\ 0\leq K\leq 1$%
, except for the Type IX universes.

The dynamical Einstein equations are 
\begin{eqnarray}
H^{\prime } &=&-(1+q)H,  \label{eq:H} \\
\Sigma _{+}^{\prime } &=&(q-2)\Sigma _{+}-S_{+}+\Pi _{+},  \label{eq:S+} \\
\Sigma _{-}^{\prime } &=&(q-2)\Sigma _{-}-S_{-}+\Pi _{-},  \label{eq:S-} \\
N_{1}^{\prime } &=&(q-4\Sigma _{+})N_{1},  \label{eq:N1} \\
N_{2}^{\prime } &=&(q+2\Sigma _{+}+2\sqrt{3}\Sigma _{-})N_{2},  \label{eq:N2}
\\
N_{3}^{\prime } &=&(q+2\Sigma _{+}-2\sqrt{3}\Sigma _{-})N_{3},  \label{eq:N3}
\end{eqnarray}%
where $q$ is deceleration parameter: 
\begin{equation}
q\equiv 2\Sigma ^{2}+\frac{1}{2}\Omega +\frac{p}{2H^{2}}=2\Sigma ^{2}+\Omega
_{\mathrm{YM}}+\frac{1}{2}(3\gamma -2)\Omega _{\mathrm{m}}\,.
\end{equation}%
The quantities $S_{\pm }$ are given only by the $N_{a}:$ 
\begin{eqnarray}
S_{+} &=&\frac{1}{6}\left\{
(N_{2}-N_{3})^{2}-N_{1}(2N_{1}-N_{2}-N_{3})\right\} , \\
S_{-} &=&\frac{1}{2\sqrt{3}}(N_{3}-N_{2})(N_{1}-N_{2}-N_{3})\,.
\end{eqnarray}%
Last, we define the Hubble-normalized anisotropic pressures of YM field, $%
\Pi _{+}$ and $\Pi _{-},$ so that 
\begin{eqnarray}
\Pi _{+} &=&-\frac{1}{6}\left[ -2\left\{ a^{\prime }+(-2\Sigma
_{+}+1)a\right\} ^{2}+\left\{ b^{\prime }+(\Sigma _{+}+\sqrt{3}\Sigma
_{-}+1)b\right\} ^{2}+\left\{ c^{\prime }+(\Sigma _{+}-\sqrt{3}\Sigma
_{-}+1)c\right\} ^{2}\right.  \nonumber \\
&&\left. -2\left( N_{1}a+{\frac{bc}{H}}\right) ^{2}+\left( N_{2}b+{\frac{ca}{%
H}}\right) ^{2}+\left( N_{3}c+{\frac{ab}{H}}\right) ^{2}\right] ,
\label{eq:Pi+} \\
\Pi _{-} &=&\frac{\sqrt{3}}{6}\left[ -\left\{ b^{\prime }+(\Sigma _{+}+\sqrt{%
3}\Sigma _{-}+1)b\right\} ^{2}+\left\{ c^{\prime }+(\Sigma _{+}-\sqrt{3}%
\Sigma _{-}+1)c\right\} ^{2}\right.  \nonumber \\
&&\left. -\left( N_{2}b+{\frac{ca}{H}}\right) ^{2}+\left( N_{3}c+{\frac{ab}{H%
}}\right) ^{2}\right] \,.  \label{eq:Pi-}
\end{eqnarray}

Here we can see the complexity introduced by the YM dynamics. 
We can see how the anisotropic pressures are driven by the directional scale
factors and their time derivatives, and not solely by the fractions of the
density as was the case with the simple model of anisotropic stresses
defined by eq. (\ref{an}).

The isotropic matter conservation equation is 
\begin{equation}
\Omega _{\mathrm{m}}^{\prime }=\{2q-(3\gamma -2)\}\Omega _{\mathrm{m}}
\label{con}
\end{equation}%
and the YM evolution equations are 
\begin{eqnarray}
a^{\prime \prime } &=&(q-2)a^{\prime }  \nonumber \\
&&+(q-1-4K+4\Sigma _{+}^{2}-4\Sigma _{+}-N_{1}(N_{2}+N_{3})+2\Pi _{+})a 
\nonumber \\
&&-\frac{1}{H}(N_{1}+N_{2}+N_{3})bc-\frac{1}{H^{2}}a(b^{2}+c^{2}),
\label{eq:a} \\
b^{\prime \prime } &=&(q-2)b^{\prime }  \nonumber \\
&&+(q-1-4K+(\Sigma _{+}+\sqrt{3}\Sigma _{-})(\Sigma _{+}+\sqrt{3}\Sigma
_{-}+2)-N_{2}(N_{3}+N_{1})-\Pi _{+}-\sqrt{3}\Pi _{-})b  \nonumber \\
&&-\frac{1}{H}(N_{1}+N_{2}+N_{3})ca-\frac{1}{H^{2}}b(c^{2}+a^{2}),
\label{eq:b} \\
c^{\prime \prime } &=&(q-2)c^{\prime }  \nonumber \\
&&+(q-1-4K+(\Sigma _{+}-\sqrt{3}\Sigma _{-})(\Sigma _{+}-\sqrt{3}\Sigma
_{-}+2)-N_{3}(N_{1}+N_{2})-\Pi _{+}+\sqrt{3}\Pi _{-})c  \nonumber \\
&&-\frac{1}{H}(N_{1}+N_{2}+N_{3})ab-\frac{1}{H^{2}}c(a^{2}+b^{2})\,.
\label{eq:c}
\end{eqnarray}

The basic variables are therefore [$H,\ \Sigma _{+},\ \Sigma _{-},\ N_{1},\
N_{2},\ N_{3},\ a,\ b,\ c$], and they are determined uniquely and
completely by Eqs. (\ref{eq:H})-(\ref{eq:N3}), (\ref{eq:a})-(\ref{eq:c}),
along with the constraint equation (\ref{eq:GFRW}).

\section{Bianchi I co-evolution of Yang-Mills fields and perfect fluids}

We compute the solutions of the dynamical equations (\ref{eq:H})-(\ref{eq:N3}) 
and (\ref{eq:a})-(\ref{eq:c}) numerically and illustrate the typical
evolutionary behaviours. In this section, we consider for simplicity only
the Bianchi type I spacetime (so $K=0$) with black-body radiation ($\gamma
=4/3$) or dust ($\gamma =1$) and a YM field. We set initial value of $H$ as $%
H_{\mathrm{ini}}=10^{-10}$ (our unit), which is not a restriction because
our time unit depends on $g_{\mathrm{YM}}$ which determines the physical
identity of the YM field and can be scaled.

First, we display the behavior of the three leading quantities, 
$\Sigma^{2},\ \Omega _{\mathrm{YM}},$ and $\Omega _{\mathrm{m}}$. 
\begin{figure}[ht]
 \includegraphics[height=6cm]{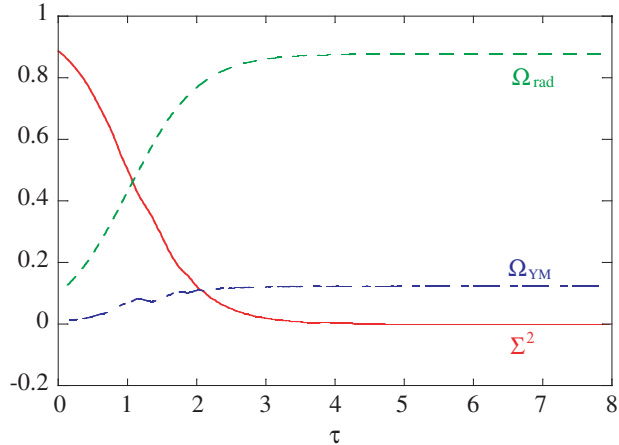} 
 \caption{The typical behaviors of $\Sigma^2,\ \Omega_{\mathrm{YM}},$ 
 and $\Omega_{\mathrm{m}}$ with increasing time. 
 This figure shows how $\Sigma^2\rightarrow 0$, and the spacetime approaches 
 the isotropic Friedmann model.
 It is remarkable that $\Omega_{\mathrm{YM}}$ does not damp.} 
 \label{fig:GFRW}
 \end{figure}
Figure \ref{fig:GFRW} shows $\Sigma ^{2}\rightarrow 0$, i.e., spacetime
approaches to the flat isotropic FRW universe. We also find that $\Omega
_{\mathrm{YM}}$ does not damp, but approaches a constant. 
It is interesting to compare this behavior with that which obtains 
close to isotropy in universes containing radiation and a magnetic field. 
There we found that $\Sigma $ and $\Omega _{\mathrm{mag}}$ both fell slowly 
as $(\ln t)^{-1}$ as $t$ increased, with asymptotic approach to an attractor 
where $\Omega_{\mathrm{mag}}/\Omega _{\mathrm{rad}}=\lambda \Sigma $, 
where $\lambda \approx O(1)$ is a calculable constant depending 
on the orientation of the field and the number of effectively massless spin states 
in the radiation background.

\begin{figure}[ht]
  \includegraphics[height=6cm]{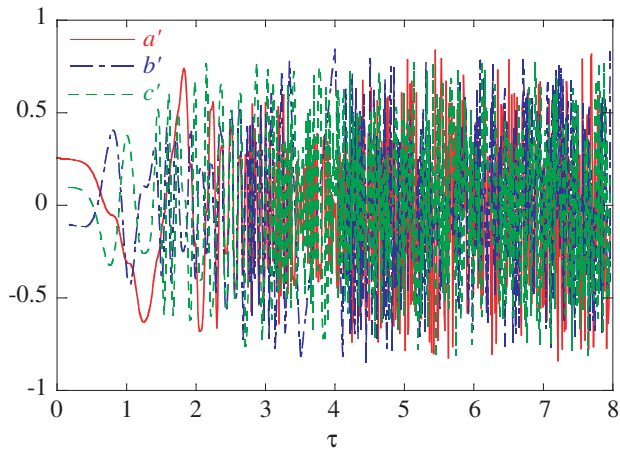} 
 \caption{Typical behaviors of the time derivative of the principal components of 
 the Yang-Mills field. The mean amplitudes of this shows amplitudes of 
 $a',\ b',$ and $c'$ remain nearly constant.} 
 \label{fig:YM'}
 \end{figure}

Second, we can display the evolution of the components of the YM field. 
 \begin{figure}[ht]
 \begin{tabular}{ccc}
  \includegraphics[height=6cm]{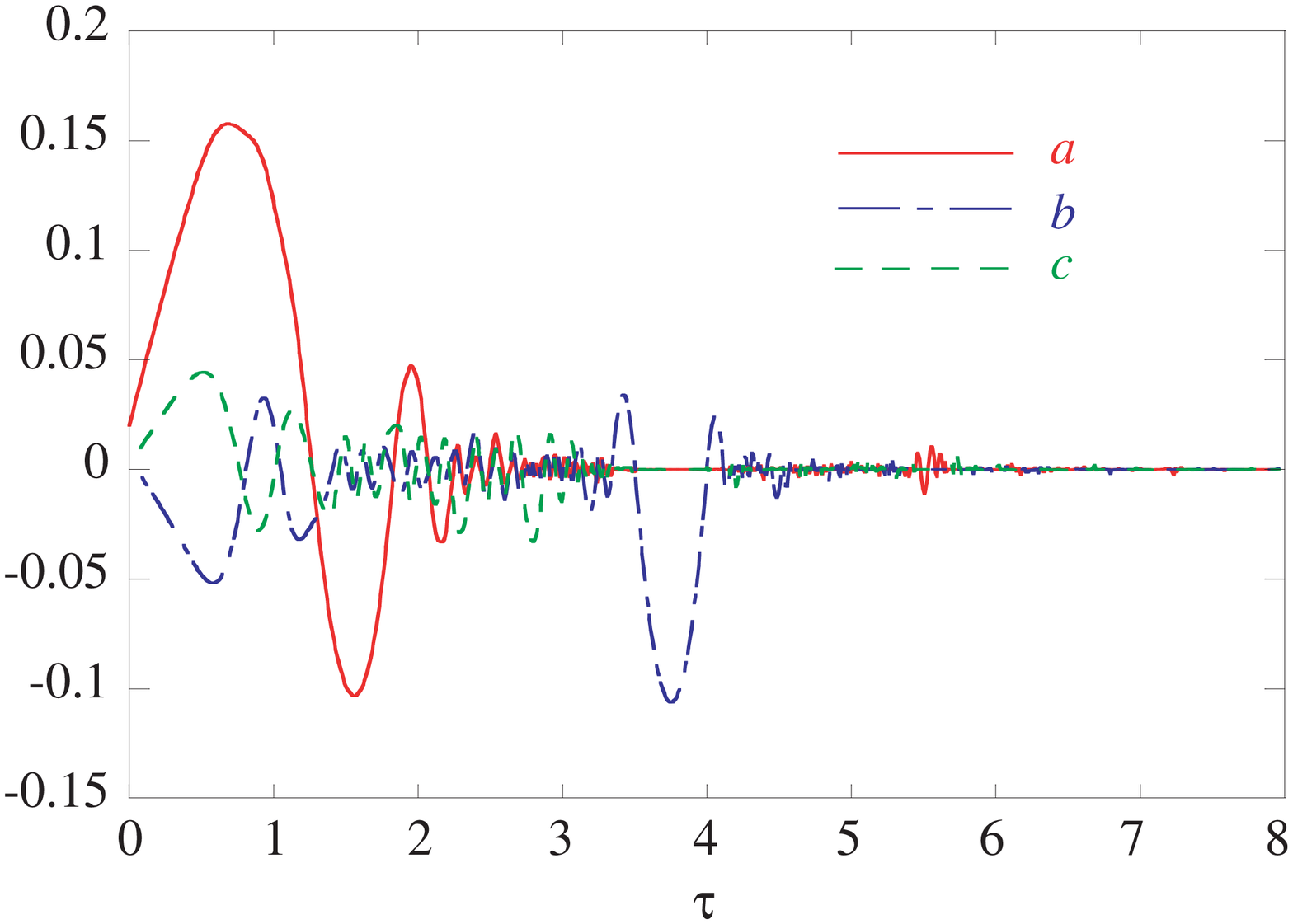} &
  \includegraphics[height=6cm]{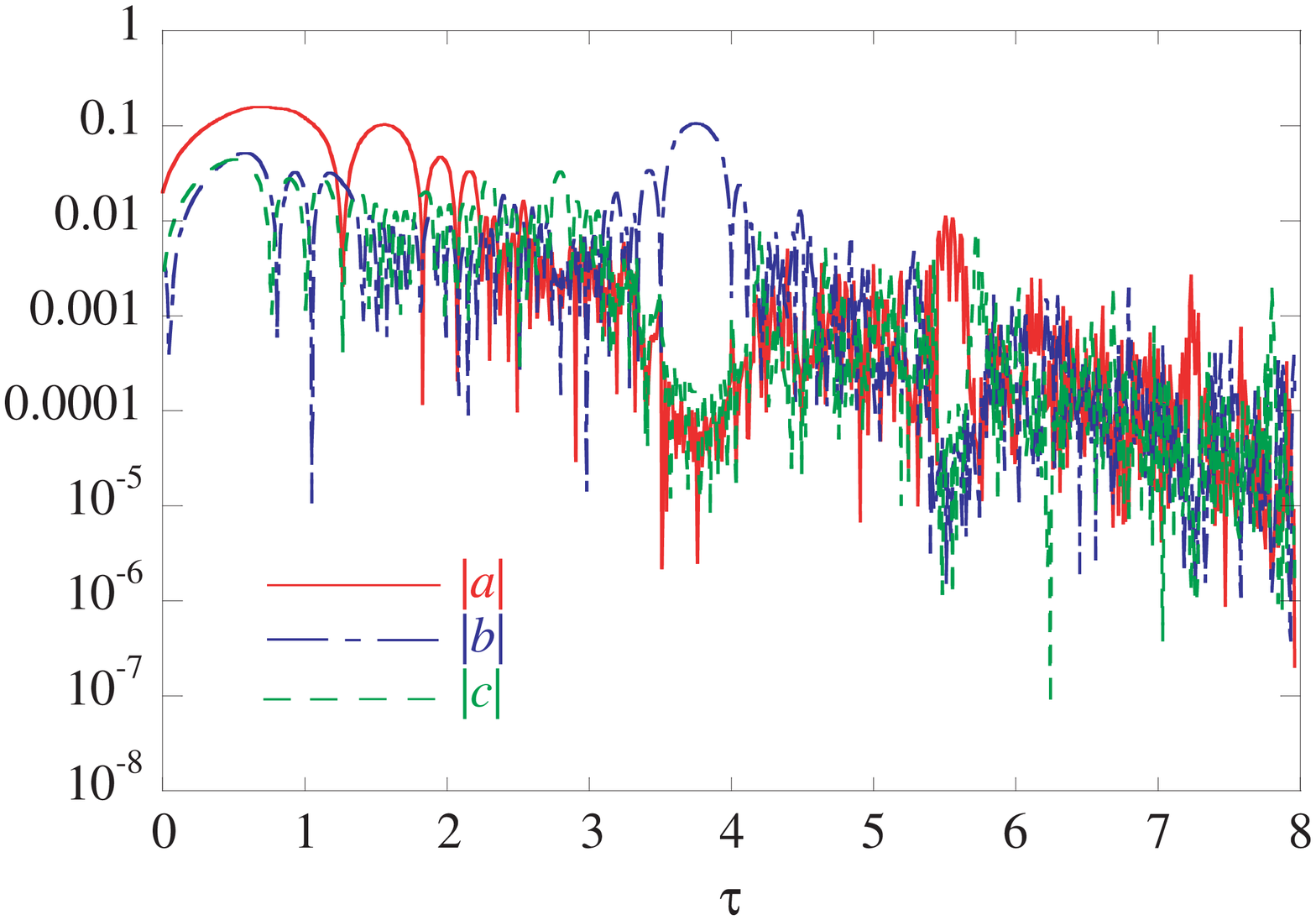} \\
  (a)&(b)\\
 \end{tabular}
 \caption{Typical behaviors of the Yang-Mills field. \\
 (a) $a,\ b,$ and $c$ damp whilst undergoing chaotic oscillations of increasing frequency. \\
 (b) The averaged damping rate of $a,\ b,$ and $c$ has the approximate form $ e^{-\tau} $. } 
 \label{fig:YM}
 \end{figure}
Figure \ref{fig:YM} shows that $a,\ b,$ and $c$ follow a sequence of damped
chaotic oscillations, and decay as $e^{-\tau }$. Figure \ref{fig:YM} also
shows that the frequencies of the oscillations of $\ a,\ b,$ and $c$ are
growing, although their amplitudes are decaying. This is further supported
by Figure \ref{fig:YM'}, which shows that the mean amplitudes of $a^{\prime },\ 
b^{\prime },$ and $c^{\prime }$ remain nearly constant. 

Third, we show the behavior of $\Pi _{\pm }$, the Hubble-normalized
anisotropic pressures. 
 \begin{figure}[ht]
  \includegraphics[height=6cm]{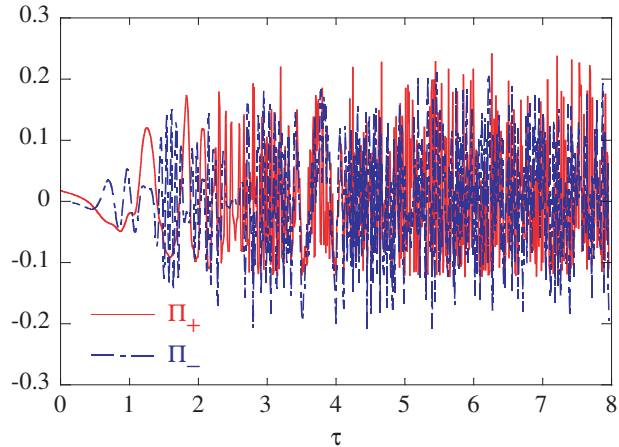} 
 \caption{The typical behaviors of the anisotropic pressure components $\Pi_{\pm}$. 
 This shows that $\Pi_{\pm}$ oscillate in a complicated way with nearly constant mean amplitudes.} 
 \label{fig:Pi}
 \end{figure}
Figure \ref{fig:Pi} shows that the $\Pi _{\pm }$ oscillate in a complicated
fashion with approximately constant mean amplitudes. This is consistent with the
observation that $\Omega _{\mathrm{YM}}\propto \Omega _{\mathrm{m}}$ but it is surprising that
the pressure anisotropy remains constant despite the fall in the shear.
These chaotic oscillations are a familiar feature of the YM evolution which
resembles the chaotic Hamiltonian dynamics of a point bouncing inside a
potential whose four steep walls in the $x-y$ plane are formed by the
branches of the rectangular hyperbolae $x^{2}y^{2}=$ constant 
\cite{blev,jin}.

In the next section we try to understand these features
emerging from the numerical studies of Bianchi I spacetime with the YM field
and radiation or dust. \emph{\ }

\subsection{The approach to isotropy}

In this section we consider Bianchi I spacetime again, i.e., $N_{a}=0$ and $K=0$. 
We assume that the FRW limit holds, that is $\Sigma \rightarrow 0,$ so
the dynamics can be close to isotropy. This results in 
\begin{eqnarray}
q = \Omega _{\mathrm{YM}}+\frac{1}{2}(3\gamma -2)\Omega _{\mathrm{m}}\,,  \label{eq:q} \\
\Omega _{\mathrm{YM}}+\Omega _{\mathrm{m}} = 1\,.  \label{eq:FRW}
\end{eqnarray}%
These conditions imply 
\begin{equation}
q=1+\frac{1}{2}(3\gamma -4)\Omega _{\mathrm{m}}\,.  \label{eq:q2}
\end{equation}%
Motivated by the results of the numerical analysis, we also assume that 
$\Omega _{\mathrm{YM}} \rightarrow \mbox{const.}$ and 
$\Omega _{\mathrm{m}} \rightarrow \mbox{const.}$ 
These assumptions reduce (\ref{con}) to the condition that 
\begin{equation}
\Omega _{\mathrm{m}}^{\prime }=\{2q-(3\gamma -2)\}\Omega _{\mathrm{m}}=0\,,
\end{equation}%
and if $\Omega _{\mathrm{m}} \neq 0$, we have 
\begin{equation}
q=\frac{1}{2}(3\gamma -2)\,.  \label{eq:q_sol}
\end{equation}%
Hence the evolution of $H$ is given by $H^{\prime }=-(1+q)H=-\frac{3}{2}%
\gamma H$. Solving this equation, we have 
\begin{equation}
H \propto \exp \left(-\frac{3}{2}\gamma \tau \right)\,.  \label{eq:expansion}
\end{equation}%
This means that\ to leading order the YM field does not determine the
expansion rate in FRW limit.

From eqs. (\ref{eq:q2}) and (\ref{eq:q_sol}), we get 
\begin{equation}
3\gamma -4=(3\gamma -4)\Omega _{\mathrm{m}}\,.
\end{equation}%
Therefore, in the radiation case $(\gamma =4/3)$, $\Omega _{\mathrm{m}}$ can converge
to any value. Otherwise $(\gamma \neq 4/3)$, $\Omega _{\mathrm{m}}$ becomes
dominant, and we must have $\Omega _{\mathrm{m}} \rightarrow 1$. 
In the radiation case ($\gamma =4/3$), the results become 
\[
q=1,\ \  \Omega _{\mathrm{YM}}+\Omega _{\mathrm{m}}=1,\ \ H \propto e^{-2\tau }\,. 
\]%
Hereafter, we will consider only the radiation case.

From our numerical results, we know that the YM field oscillates with
growing frequency while its amplitude is damped. Therefore, we
assume that $a\rightarrow 0,a^{\prime }\not\rightarrow 0$. Because the $\Pi
_{\pm }$ are both finite\footnote{%
In Bianchi I, II, and VI$_{0}$ spacetimes, $-1\leq \Pi _{+}\leq 2$ and 
$-\sqrt{3}\leq \Pi _{-}\leq \sqrt{3}$ (see \cite{jin}).}, 
eq. (\ref{eq:a}) becomes 
\begin{equation}
a^{\prime \prime }=-a^{\prime }-\frac{1}{H^{2}}a(b^{2}+c^{2})=-a^{\prime
}-e^{4\tau }a(b^{2}+c^{2})\,.  \label{eq:a_2}
\end{equation}%
If we divide $a$ into an amplitude and a phase, with $a=e^{-p\tau
}e^{i\theta _{1}}$, where $p$ is the damping rate and $\theta _{1}(\tau )$
is the time-dependent oscillation frequency, then 
\begin{equation}
a^{\prime }=-pa+i\theta _{1}^{\prime }a=(-p+i\theta _{1}^{\prime })a
\label{A}
\end{equation}%
In order to keep $a^{\prime }\not\rightarrow 0$, $\theta _{1}^{\prime }$ has
to grow like 
\[
\theta _{1}^{\prime }\sim O(e^{p\tau })\,. 
\]%
Differentiating eq. (\ref{A}), this gives 
\begin{eqnarray}
a^{\prime \prime } &=&-pa^{\prime }+i\theta _{1}^{\prime \prime }a+i\theta
_{1}^{\prime }a^{\prime }  \nonumber \\
&=&\{p^{2}-\theta _{1}^{\prime }{}^{2}+i(\theta _{1}^{\prime \prime
}-2p\theta _{1}^{\prime })\}a\,,  \label{eq:a''}
\end{eqnarray}%
where $\theta _{1}^{\prime }{}^{2}$ grows $O(e^{2p\tau })$ and $\theta
_{1}^{\prime \prime }$ depends on the phase of $\theta _{1}^{\prime }$.
Substituting in eq. (\ref{eq:a_2}) yields 
\begin{eqnarray}
p^{2}-\theta _{1}^{\prime }{}^{2}+i(\theta _{1}^{\prime \prime }-p\theta
_{1}^{\prime }) &=&-(-p+i\theta _{1}^{\prime })-e^{4\tau }(b^{2}+c^{2}) 
\nonumber \\
&=&p-i\theta _{1}^{\prime }-e^{4\tau }e^{-2p\tau }(e^{2i\theta
_{2}}+e^{2i\theta _{3}})\,,  \label{eq:a_3}
\end{eqnarray}%
where, for the simplicity, we assume $b$ and $c$ damp at the same rate as $a$%
, so $b=e^{-p\tau }e^{i\theta _{2}}$ and $c=e^{-p\tau }e^{i\theta _{3}}$.
Picking out the dominant terms in eq. (\ref{eq:a_3}), we have 
\begin{eqnarray}
-\theta _{1}^{\prime }{}^{2}+i\theta _{1}^{\prime \prime } &=&-e^{(4-2p)\tau
}(e^{2i\theta _{2}}+e^{2i\theta _{3}})  \nonumber \\
&=&-e^{(4-2p)\tau }2\cos (\theta _{2}-\theta _{3})e^{i(\theta _{2}+\theta
_{3})}\,.
\end{eqnarray}%
It is generic to assume that there is no relation between $\theta _{2}$ and $%
\theta _{3}$, so $\theta _{2}\neq \theta _{3}$. As a result, the
characteristic frequencies of the above equation will be neither purely 
real nor purely imaginary. Therefore it is likely that $O(\theta _{1}^{\prime
}{}^{2})\sim O(\theta _{1}^{\prime \prime })$, which means that the
leading-order solution of the above equation grows\emph{\ }as $e^{2p\tau }$,
i.e., $p=1$.

\medskip

From this, we can estimate $\Pi _{\pm }$ as follows, 
\begin{eqnarray}
\Pi _{+} &=&-\frac{1}{6}\left( -2a^{\prime }{}^{2}+b^{\prime
}{}^{2}+c^{\prime }{}^{2}-2e^{2i(\theta _{2}+\theta _{3})\tau }+e^{2i(\theta
_{3}+\theta _{1})\tau }+e^{2i(\theta _{1}+\theta _{2})\tau }\right) \sim O(%
\mbox{const.})  \nonumber \\
\Pi _{-} &=&\frac{\sqrt{3}}{6}\left( -b^{\prime }{}^{2}+c^{\prime
}{}^{2}-e^{2i(\theta _{3}+\theta _{1})\tau }+e^{2i(\theta _{1}+\theta
_{2})\tau }\right) \sim O(\mbox{const.})\,.
\end{eqnarray}%
It follows from our discussion that $\Pi _{\pm }\not\rightarrow 0$ despite
the fact that$\ \Sigma \rightarrow 0\,.$ This result is quite unexpected.

\begin{figure}[ht]
 \includegraphics[height=6cm]{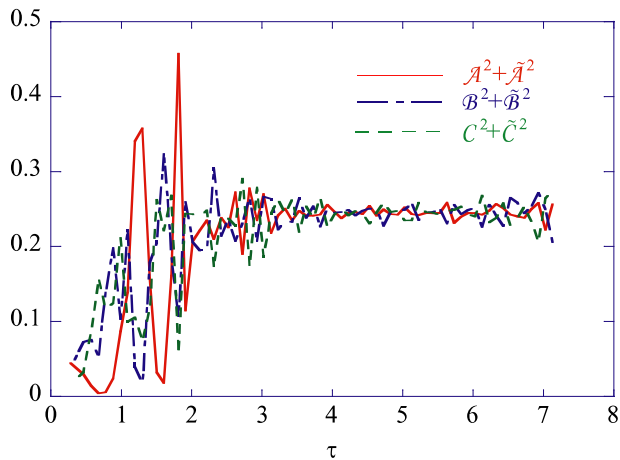} 
 \caption{The typical behaviors of the 
 $\langle\mathcal{A}^2 + \tilde{\mathcal{A}}^2\rangle$, 
 $\langle\mathcal{B}^2 + \tilde{\mathcal{B}}^2\rangle$, and 
 $\langle\mathcal{C}^2 + \tilde{\mathcal{C}}^2\rangle$, 
 where $\langle \cdots \rangle$ means time average per $0.1\tau$. 
 This shows that $\mathcal{A}^2 + \tilde{\mathcal{A}}^2$, 
 $\mathcal{B}^2 + \tilde{\mathcal{B}}^2$, and $\mathcal{C}^2 + \tilde{\mathcal{C}}^2$ 
 are nearly equal after the shear is significantly reduced, i.e. $\tau>2$, 
 although $\mathcal{A}^2,\ \mathcal{B}^2,\ \mathcal{C}^2,\ 
 \tilde{\mathcal{A}}^2,\ \tilde{\mathcal{B}}^2,$ and $\tilde{\mathcal{C}}^2$ 
 oscillate strongly.} 
 \label{fig:ABC}
 \end{figure}

It is worth checking explicitly that these non-zero anisotropic pressures, $%
\Pi _{\pm },$ do not break isotropy. That is, we need to confirm whether
these non-zero $\Pi _{\pm }$ make $\Sigma \not\rightarrow 0$ or not. Using
eqs. (\ref{eq:Pi+}) and (\ref{eq:Pi-}), and the property that $\Sigma _{\pm
}\ll \Pi _{\pm }$, and assuming 
$\mathcal{A}^2 + \tilde{\mathcal{A}}^2 \sim 
\mathcal{B}^2 + \tilde{\mathcal{B}}^2 \sim 
\mathcal{C}^2 + \tilde{\mathcal{C}}^2$,
where $\mathcal{A}\equiv a^{\prime }+(-2\Sigma _{+}+1)a$, 
$\mathcal{B}\equiv b^{\prime }+(\Sigma _{+} + \sqrt{3}\Sigma _{-}+1)b$, 
$\mathcal{C}\equiv c^{\prime }+(\Sigma _{+} - \sqrt{3}\Sigma _{-}+1)c$, 
$\tilde{\mathcal{A}}\equiv N_{1}a+\frac{bc}{H}$, 
$\tilde{\mathcal{B}}\equiv N_{2}b+\frac{ca}{H}$, and 
$\tilde{\mathcal{C}}\equiv N_{3}c+\frac{ab}{H}$, we have 
\begin{eqnarray}
\Sigma ^{\prime } &=&\frac{(\Sigma ^{2})^{\prime }}{2\Sigma }=\frac{1}{%
\Sigma }(\Sigma _{+}\Sigma _{+}^{\prime }+\Sigma _{-}\Sigma _{-}^{\prime
})\sim \frac{1}{\Sigma }(\Sigma _{+}\Pi _{+}+\Sigma _{-}\Pi _{-})  \nonumber
\\
&=&\frac{1}{3}\left[ (\mathcal{A}^{2}+\tilde{\mathcal{A}}^{2})\cos \Psi +(%
\mathcal{B}^{2}+\tilde{\mathcal{B}}^{2})\cos \left( \Psi +\frac{2}{3}\pi
\right) +(\mathcal{C}^{2}+\tilde{\mathcal{C}}^{2})\cos \left( \Psi -\frac{2}{%
3}\pi \right) \right]  \nonumber \\
&\sim &\frac{1}{3}\left[ (\mathcal{A}^{2}+\tilde{\mathcal{A}}^{2})\left(
\cos \Psi +\cos \left( \Psi +\frac{2}{3}\pi \right) +\cos \left( \Psi -\frac{%
2}{3}\pi \right) \right) \right] =0\,,
\end{eqnarray}%
where $\Psi \equiv \tan ^{-1}(\Sigma _{-}/\Sigma _{+})\,.$ This shows that
the non-zero $\Pi _{\pm }$ modes do \textit{not} break isotropy. 
The assumption, $\mathcal{A}^2 + \tilde{\mathcal{A}}^2 \sim 
\mathcal{B}^2 + \tilde{\mathcal{B}}^2 \sim 
\mathcal{C}^2 + \tilde{\mathcal{C}}^2$, is justified by numerical solutions. 
Figure \ref{fig:ABC} shows the time averages of $\mathcal{A}^2 + \tilde{\mathcal{A}}^2 \sim 
\mathcal{B}^2 + \tilde{\mathcal{B}}^2 \sim \mathcal{C}^2 + \tilde{\mathcal{C}}^2$ 
over a period of $0.1 \tau$, which confirms our assumption. 

\medskip In summary: so far our analyses have shown that if we assume that
the radiation-YM universe approaches FRW, in the sense that $\Sigma
\rightarrow 0,$ and also that $\Omega _{\mathrm{YM}} \rightarrow \mbox{const.}$ 
and $\Omega _{\mathrm{m}} \rightarrow \mbox{const.},$ then the YM field components 
satisfy $a\rightarrow 0$, and $a^{\prime} \rightarrow O(\mbox{const.}),$ 
with $\mathcal{A}^2 + \tilde{\mathcal{A}}^2 \sim 
\mathcal{B}^2 + \tilde{\mathcal{B}}^2 \sim 
\mathcal{C}^2 + \tilde{\mathcal{C}}^2$, 
and it is found that $H\propto e^{-2\tau },$ $a\propto
O(e^{-\tau }),$ $\Sigma ^{\prime }\rightarrow 0$, and the normalised
pressures $\Pi _{\pm }$ oscillate with nearly constant amplitudes. In the
next section we provide some further details of this evolution which can be
obtained from the numerical studies.

\section{A survey of typical evolutionary behaviors}

Our numerical studies of the Bianchi I expansion dynamics were performed for
various combinations of radiation or dust and YM field. We give here a
summary of the principal conclusions from these investigations. These
results are stable against small changes in the initial data ($H_{\mathrm{ini}}$ 
and $\Sigma _{\mathrm{ini}}$) and appear to be robust. This robustness
is a consequence of the generic nature of the chaotic oscillations of the YM
field and the fact that these chaotic oscillations have a small effect
on the expansion dynamics of the universe.

\subsection{ Radiation and YM field}

\subsubsection{YM field dominates: $\ \Omega _{\mathrm{YM}} \gg \Omega _{\mathrm{rad}}$}

In the case where the YM field density dominates the density of the
isotropic black-body radiation field, we find that 
$\Omega _{\mathrm{YM}} \sim \mathrm{const.}$ and $\Omega _{\mathrm{rad}}\sim $ \textrm{const. }%
during the evolution. If we define an averaged expansion scale factor for
the expanding universe, $\ell (t)$, by

\[
 H\equiv \frac{d\ell /dt}{\ell }
\]
then the normalised shear is found to fall as $\Sigma \propto \ell ^{-1.01}$
in the numerical integrations, while the Hubble expansion rate falls at the
rate expected in a FRW model, with $H\propto \ell ^{-2.00},$ so the mean
scale factor evolves as $\ell \propto t^{0.500}\sim t^{1/2}$, as in
a FRW universe.

\subsubsection{Radiation dominates: $\Omega _{\mathrm{YM}} \ll \Omega _{\mathrm{rad}}$}

If the black-body radiation density dominates the YM field then we still
find $\Omega _{\mathrm{YM}}\sim \mathrm{const.}$ and $\Omega _{\mathrm{rad}%
}\sim \mathrm{const.}$ and the normalised shear falls almost linearly with
the scale factor, as $\Sigma \propto \ell ^{-1.07},$with $H\propto \ell
^{-2.00}$ and $\ell \propto t^{0.500}\sim t^{1/2}$ as in the FRW model. Thus
we see that the evolution in both of the radiation plus YM field situations
is similar, with

\[
\Sigma \propto \ell ^{-1} \ \text{and} \ \ell \propto t^{1/2} 
\]%
holding to an excellent approximation. Although the anisotropic pressure can
be significant, it is not driving significant shear expansion anisotropy.

\subsection{Dust and YM field}

\subsubsection{YM field dominates: $\Omega _{\mathrm{YM}} \gg \Omega_{\mathrm{dust}} $}

When the YM field dominates over the dust we are in the very early stages of
the overall evolution because on average the dust density redshifts away
more slowly than the YM field. When the YM field still dominates, so $\Omega
_{\mathrm{YM}}\sim 1$, the numerical evolution gives

\begin{eqnarray*}
\Omega _{\mathrm{dust}} &\propto &\ell ^{0.983}\  \\
\Sigma &\propto &\ell ^{-1.01}
\end{eqnarray*}%
while $H\propto \ell ^{-2.00}$ so $\ell \propto t^{0.500}\sim t^{1/2}.$ This
reflects the assumption of the YM field domination and is not inconsistent
with eq. (\ref{eq:expansion}) because the assumption used there, $\Omega _{%
\mathrm{YM}}$ and $\Omega _{\mathrm{m}}\rightarrow \mathrm{const}$, is
violated. We see that the damping rate of shear and the expansion law are
the same as that in the radiation-dominated case. It is understandable that $%
\Omega _{\mathrm{dust}}\propto \ell ^{0.983}$, because $\mu _{\mathrm{dust}%
}\propto \ell ^{-3}$ and $H\propto t^{-1}\propto \ell ^{-2}$. This results
in the dependence $\Omega _{\mathrm{dust}}\equiv \mu _{\mathrm{dust}%
}/3H^{2}\propto \ell $, as expected. Thus, $\Omega _{\mathrm{dust}}$ grows
and will eventually become dominant and the assumption that $\Omega
_{\mathrm{YM}} \gg \Omega _{\mathrm{dust}}$ will eventually fail.

\subsubsection{Dust dominates: $\Omega _{\mathrm{YM}} \ll \Omega _{\mathrm{dust}}$}

This is the natural situation for the universe to evolve into after the
radiation-dominated era. Our numerical studies find that in the dust-dominated
phase the YM density falls as

\[
\Omega _{\mathrm{YM}}\propto \ell ^{-0.954} 
\]
while the normalised shear falls as

\[
\Sigma \propto \ell ^{-1.48}\sim \ell ^{-3/2} 
\]%
and the Hubble rate falls as $H\propto \ell ^{-1.52}$. Hence, we obtain a close
approximation to the evolution expected in a dust-dominated FRW universe,
with $\ell \propto t^{0.658}\sim t^{2/3}$. As expected, the shear falls off
more rapidly than in the radiation-dominated case because of the growing
influence of the isotropising dust density. It is also interesting that the
shear falls off more rapidly than in the simpler dust plus magnetic
universes \cite{Barrow}. We note that the fall of the shear in the YM case
is close to the $\Sigma \propto \ell ^{-3/2}$ fall-off that would occur if
the YM field were absent in a dust-dominated Bianchi I universe. We see how
this arises by noting that

\[
\Sigma _{+}^{\prime }=-\frac{3}{2}\Sigma _{+}+\Pi _{+}\sim -\frac{3}{2}%
\Sigma _{+} 
\]%
because $|\Pi _{\pm }|<\Omega _{\mathrm{YM}}$, and therefore,

\[
\Sigma _{+}\sim \exp \left(-\frac{3}{2}\tau \right)\sim \ell ^{-3/2}. 
\]

\section{Comparison of Yang-Mills and magnetic fields}

One of the original motivations for our study of the evolution of YM fields
in the presence of perfect fluids was the unusual behavior found in the
case of a pure magnetic field and a perfect fluid, notably in the situation
where the perfect fluid is black-body radiation. Since the YM field is a
generalisation of a magnetic field it is instructive to compare and contrast
the results for these two cases. \emph{\ }

In the case of an almost isotropic universe ($\ell \propto t^{1/2}$)
containing a pure magnetic field (or other anisotropic stresses with
pressure anisotropy of the form (\ref{an})) and black-body radiation with 
$\Omega _{\mathrm{mag}}\ll \Omega _{\mathrm{rad}}$, the expansion-normalised
shear falls logarithmically in time, and

\[
\Sigma \propto \mu _{\mathrm{mag}}/\mu _{\mathrm{rad}}\propto 1/\ln t\propto
1/\ln \ell 
\]%
This unusual 'critical' evolution arises because there is a zero eigenvalue
when we linearise the Bianchi type I magnetic radiation universe around
the isotropic Friedmann radiation universe \cite{collins, Barrow}. Note also
that the ratio of the magnetic to the black-body density is not constant as
is often assumed, but falls slowly due to the coupling to the shear \cite%
{zeld}.

There is a late-time attractor with $\mu _{\mathrm{mag}}/\mu _{\mathrm{rad}%
}\approx \Sigma $ and $\ $%
\[
\Omega _{\mathrm{mag}}\equiv \mu _{\mathrm{mag}}/3H^{2}\propto \mu _{\mathrm{%
mag}}/\mu _{\mathrm{rad}}\propto \Sigma . 
\]%
It is interesting that $\Omega _{\mathrm{mag}}\propto \Sigma $. This means
that $\Omega _{\mathrm{mag}}$ can be constrained directly by the CMB
temperature anisotropy, $\Delta T/T$, since the presently observed $\Delta
T/T\sim \Sigma _{\mathrm{rec}}$ where $z_{\mathrm{rec}}\sim 1100$ is the recombination
redshift. After accounting for the short period of dust dominated evolution
from the equal-density redshift, $z_{\mathrm{eq}}\sim 10^{4}$, to $z_{\mathrm{rec}}$ this
leads to strong limits on any spatially homogeneous cosmological magnetic
field today of $3.4\times 10^{-9}(\Omega _{0}h_{50}^{2})^{1/2}G$, \cite{maglim}, 
or on any anisotropic stress with pressures of the form (\ref{an}),
see ref. \cite{Barrow} for details and examples.

In a dust-dominated era ($\Omega _{\mathrm{mag}}\ll \Omega _{\mathrm{dust}}$%
) the magnetic stresses still slow the fall off of the shear to $\Sigma
\propto \ell ^{-1}\propto t^{-2/3},$ whereas we would have had $\Sigma
\propto \ell ^{-3/2}$ if the magnetic field was absent. The magnetic density
evolves as

\[
\Omega _{\mathrm{mag}}\propto \mu _{\mathrm{mag}}/\mu _{\mathrm{dust}%
}\propto \ell ^{-1}. 
\]

The results for the different magnetic and YM evolutions are summarised in
Table \ref{table1}.

\begin{table}[h]
\begin{tabular}{c||c|c}
Material content & Shear Evolution, $\Sigma $ & \ Anisotropic stress density \ 
\\ \hline \hline
Dust + magnetic field & $\Sigma \propto \ell ^{-1}\propto t^{-2/3}$ & 
$\Omega _{\mathrm{mag}}\propto \Sigma \propto t^{-2/3}$ \\ \hline
Dust + YM field & $\Sigma \propto \ell ^{-3/2}\propto t^{-1}$ & 
$\Omega_{\mathrm{YM}}\propto t^{-2/3}$ \\ \hline
\ Radiation + magnetic field \ & 
\ $\Sigma \propto (\ln \ell )^{-1}\propto (\ln t)^{-1}$ \ & 
$\Omega _{\mathrm{mag}}\propto \Sigma \propto (\ln t)^{-1}$ \\ \hline
Radiation + YM field & $\Sigma \propto \ell ^{-1}\propto t^{-1/2}$ & 
$\Omega_{\mathrm{YM}}=$ const.%
\end{tabular}
\caption{Evolution of the normalised shear, $\Sigma \equiv \sigma /H$.} \label{table1}
\end{table}

\section{Observational bounds}

\subsection{The microwave background }

The key difference that these results display between the magnetic and YM
fields is the rapid fall-off in the shear that accompanies the YM evolution.
As a result the YM field density does not determine a shear attractor at the
end of the radiation era and does not have a significant effect on the CMB
anisotropy in the way that the magnetic field does. As a corollary,
observations of the CMB temperature anisotropy do not provide a direct and
powerful upper bound on the present YM field density in the way that they do
for a magnetic field density. A scenario with which $\mu _{\mathrm{YM}}\sim
\mu _{\mathrm{rad}}$ is not constrained by shear anisotropy of CMB,
although it may be constrained by Big Bang nucleosynthesis (BBN), as we
discuss below.

The slow decay of the shear in the magnetic universe means that the CMB
anisotropy places a stronger limit on the magnetic field density than can be
obtained by considering its effects on the expansion rate of the universe at
the time of BBN, $z_{\mathrm{BBN}}\sim 10^{9}-10^{10}$. 
However, in the YM case the rapid fall in the shear 
as we go forward in time means that a negligible
shear at the the time of recombination will be much larger at earlier
epochs, with an enhancement factor of

\[
\frac{\Sigma _{\mathrm{BBN}}}{\Sigma _{\mathrm{rec}}}\sim 
\frac{1+z_{\mathrm{BBN}}}{1+z_{\mathrm{eq}}}\sim 10^{5}.
\]%
Since $\Sigma _{\mathrm{BBN}}<0.35$ from the helium-4 abundance \cite%
{wainwright-ellis, JB, JB2, JB3, sw}, we see that the BBN anisotropy bound
is 
\begin{equation}
\Sigma _{\mathrm{rec}}\lesssim 3.5\times 10^{-6}.  \label{bbn}
\end{equation}%
This is about an order of magnitude stronger than the limit imposed by the
CMB since \cite{maglim}%
\begin{equation}
\Delta T/T=\Sigma _{\mathrm{rec}}\times f(\theta ,\phi ,\Omega _{0}),
\end{equation}%
where $f\sim O(1)$ is a dimensionless pattern orientation factor, which is
bounded as $0.6<f<2.2$ in flat or open universe by the COBE results over $%
\theta >10^{0}$ angular scales and so the Bianchi I CMB constraint on $%
\Sigma _{\mathrm{rec}}$ is only 
\begin{equation}
\Sigma _{\mathrm{rec}}<10^{-5}\,.  \label{cmb}
\end{equation}%
As discussed in Barrow \cite{pep}, this simple anisotropy evolution leads to
an unphysical super-Planck anisotropy energy density in the shear modes at
very early times if extrapolated backwards to $z \gg z_{\mathrm{eq}}$ from modest values
of shear at $z_{\mathrm{eq}}$. A physically reasonable requirement (the 'Planck
Equipartition Principle (PEP)' would be that at $t_{pl}\sim G^{1/2}\sim
10^{-43}s$ all energy densities contributing to the cosmological dynamics
should be bounded above by the Planck density $\mu _{pl}\sim t_{pl}^{-4}\sim
10^{94}gm.cm^{-3}$. This would require $\Sigma _{pl}\leq 1$ at \ $z_{pl}\sim
10^{32}$ and hence

\[
\frac{\Sigma _{\mathrm{rec}}}{\Sigma _{pl}}\sim \frac{1+z_{\mathrm{rec}}}{1+z_{pl}} 
\sim 10^{-29}
\]%
and the residual anisotropy in the CMB on large angular scales in the YM
Bianchi I universe would be completely negligible. By contrast, in the
magnetic universe the slow logarithmic fall in $\Sigma $ allows an
interesting anisotropy level $\Sigma _{\mathrm{rec}}\sim O(10^{-5})$ to persist at
recombination even if $\Sigma _{pl}\leq 1.$ The predicted value of $\Delta
T/T\sim \Sigma _{\mathrm{rec}}$ depends upon the number of relativistic spin states 
contributing to the equilibrium radiation sea and can also differ 
if other forms of anisotropy are assumed \cite{pep}. In general, we
would expect that YM fields in the more complex Bianchi type VII universes
would display slow logarithmic decay of their shear regardless of the
presence or absence of the YM fields. This is due to the anisotropic
3-curvature which mimics the presence of an anisotropic 'fluid' of
long-wavelength gravitational waves satisfying (\ref{an}) to a good
approximation and leads to $\Sigma \propto (\ln t)^{-1}$, see refs. \cite%
{DLN, lukash, BJS, BS, Jaffe}. Note however, that these modes are excluded
in anisotropic open universes if their spatial topology is compact \cite%
{bkodama1,bkodama2}: hyperbolic Bianchi spaces with compact topology must be
isotropic.

\subsection{Constraints from BBN}

In the scenario containing YM fields studied above, the YM density $\Omega _{%
\mathrm{YM}}$ does not evolve in proportional to the shear. The YM pressure
anisotropy does not dominate the simple 'adiabatic' decay of the shear ($%
\sigma \propto \ell ^{-3}$) that occurs in the absence of anisotropic
sources or anisotropic 3-curvature. Therefore it is impossible to constrain $%
\Omega _{\mathrm{YM}}$ by the shear anisotropy alone, in the way that $%
\Omega _{\mathrm{mag}}$ was constrained by shear. Primordial nucleosynthesis
gives the strongest constraint on $\Omega _{\mathrm{YM}}$. In our YM
scenario, the expansion rate $H$ is larger than in the isotropic FRW case,
because

\[
H^{2}=(8\pi G/3)(\mu _{\mathrm{rad}}+\mu _{\mathrm{YM}})
\]%
where $\mu _{\mathrm{rad}}$ $\propto T^{4}$ is fixed by observation. The
increase in the expansion rate raises the temperature at which
neutron-proton abundance ratio freezes out of equilibrium, and this leads to
an enhancement in the final helium-4 abundance. The YM field evolves on
average like a radiation field and we can use constraints on the density of
dark radiation (DR) \cite{Ichiki} to limit its allowed effect in this
process. This gave a bound of $\Omega _{\mathrm{DR_{BBN}}}<0.105\Omega _{%
\mathrm{rad_{BBN}}}$ at $2\sigma $ confidence level. Therefore, we expect
the YM field density to be similarly constrained with the bound

\[
\Omega _{\mathrm{YM}}<0.105\Omega _{\mathrm{rad}}.
\]

A similar bound would be obtained from BBN considerations for the magnetic
energy density. However, the bound previously obtained from the CMB
anisotropy is far stronger in this case: $\Omega _{\mathrm{mag}} \lesssim
10^{-5}\Omega _{\mathrm{rad}}.$

\section{Conclusions}

We have analysed the evolution of YM fields in anisotropic radiation and
dust cosmologies which evolve close to isotropy. The YM field undergoes
chaotic oscillations which produce small chaotic vibrations about the
Friedmann expansion when the anisotropy level is small. We investigated the
evolution of the shear anisotropy and the YM density during the radiation
and dust eras. This revealed significant differences to the unusual
situation that is known to exist in magnetic cosmologies containing perfect
fluids. In particular, unlike in magnetic universes, there is no attractor
in the radiation era which couple the YM energy density to the shear
anisotropy. Consequently, there is no direct bound on the YM field density
from the CMB temperature anisotropy. We have carried out a comparative
analysis of the magnetic and YM universes which shows how magnetic universes
of Bianchi type I are principally constrained by the CMB anisotropy whilst
the YM universes of this type are constrained by the effects of the YM
energy density on the synthesis of helium-4. The YM evolution reveals
unusual features. Despite the fall in shear to Hubble expansion ratio, the
Hubble-normalised pressure anisotropies induced by the YM field do not decay
during the radiation era, but oscillate chaotically with constant
amplitudes. 

\section*{ACKNOWLEDGMENTS}
This work was partially supported by a Grant for The 21st Century COE Program
(Holistic Research and Education Center for Physics Self-organization Systems)
at Waseda University.

\end{document}